\documentclass[aps,prl,10pt,twocolumn,showpacs,superscriptaddress]{revtex4}

\usepackage{graphicx}
\usepackage{amssymb}
\usepackage{amsmath}
\usepackage{color}

\def\ket#1{  \left\vert  #1   \right\rangle   }
\def\bra#1{  \left\langle  #1   \right\vert   }
\def\mem#1#2#3{  \left\langle #1 \left\vert  #2 \right\vert #3 \right\rangle   }
\def\twobytwo#1#2#3#4{  \left( \! \begin{array}{cc}
                                   #1 & #2   \\[0.2cm]
                                   #3 & #4   \end{array} \right) \! }

\newcommand{\etal}{{\it et al.}}

\begin{document}
%
% ------------------------------------------ Title ------------------------------------------------
%
\title{Radiative Capture of Twisted Electrons by Bare Ions}

%
% -------------------------------------- Authors and affiliations ----------------------------------
%
\author{O.~Matula}
\affiliation{Physikalisches Institut, Ruprecht--Karls--Universit\"at Heidelberg, D--69120 Heidelberg, Germany}

\author{A.~G.~Hayrapetyan}
\affiliation{Physikalisches Institut, Ruprecht--Karls--Universit\"at Heidelberg, D--69120 Heidelberg, Germany}

\author{V.~G.~Serbo}
\affiliation{Novosibirsk State University, RUS--630090 Novosibirsk, Russia}

\author{A.~Surzhykov}
\affiliation{Helmholtz--Institut Jena, D--07743 Jena, Germany}

\author{S.~Fritzsche}
\affiliation{Helmholtz--Institut Jena, D--07743 Jena, Germany}
\affiliation{Theoretisch--Physikalisches Institut, Friedrich--Schiller--Universit\"at Jena, D–-07743 Jena, Germany}

\date{\today}

%
%
% ------------------------------------------------- Abstract --------------------------------------------
%
%
\begin{abstract}
Recent advances in the production of twisted electron beams with a subnanometer spot size offer unique opportunities to explore the role of orbital angular momentum (OAM) in basic atomic processes. 
In the present work, we address one of these processes: radiative recombination of twisted electrons with bare ions. 
Based on the density matrix formalism and the non-relativistic Schr\"odinger theory, analytical expressions are derived for the angular distribution and the linear polarization of photons emitted due to the capture of twisted electrons into the ground state of (hydrogen-like) ions. 
We show that these angular and polarization distributions are sensitive to both, the transverse momentum and the topological charge of the electron beam. 
To observe in particular the value of this charge, we propose an experiment that makes use of the coherent superposition of two twisted beams.
\end{abstract}

\maketitle

%
%
% ------------------------------------------------- Introduction --------------------------------------------
%
%

\textit{Introduction}--- Beams of electrons that carry each a well-defined projection $\hbar m$ of orbital angular momentum (OAM) onto the propagation direction currently attract considerable interest in both fundamental and applied research \cite{Bliokh2007,Uchida2010,Verbeeck2010,McMorran2011}. 
These so-called \textit{twisted} or \textit{vortex} beams can be readily generated with a kinetic energy of up to a few hundred keV and with a topological charge $m$ as high as $m=100$ \cite{Uchida2010,Verbeeck2010,McMorran2011,Verbeeck2012,Saitoh2012}. 
Their unique features make these beams an ideal tool for novel studies in various fields of modern physics. 
Scattering of twisted electrons on a solid target, for example, can provide detailed insights into the magnetic structure of the target materials \cite{Verbeeck2010,Lloyd2012,Saitoh2013}. 
When propagating in external (electric and magnetic) fields, beams with non-zero OAM help explore the vacuum Faraday effect \cite{Greenshields2012}, Larmor and Gouy rotations \cite{Guzzinati2013}, or the formation of modified Landau and Aharonov-Bohm levels \cite{Gallatin2012, Bliokh2012}. 
Moreover, the production of electron vortex beams with a subnanometer focal spot size \cite{Verbeeck2011,Verbeeck2012} opens up new possibilities to investigate effects originating from the transfer of OAM in \textit{electron-ion collisions}.

One fundamental process that takes place when electrons collide with matter is the capture of an electron into an atomic or ionic bound state accompanied by the emission of a photon \cite{Eichler2007}. 
Such a \textit{radiative recombination} (RR), which can be viewed as the time-reversed photoionization, occurs frequently in stellar and laboratory plasmas as well as in ion trap and storage ring experiments. 
During the last decades, therefore, this process has attracted much attention both in experiment and theory \cite{Scofield1989,Stoehlker1995,Tashenov2006}. 
Even though these investigations have provided valuable knowledge on spin, relativistic, and quantum electrodynamic (QED) effects in the electron-photon coupling, they were restricted to the case of ``standard'' plane-wave electron beams---beams without an OAM along the propagation axis.

Twisted electron beams offer a new and intriguing degree of freedom for the RR process: a {\it non-zero} orbital momentum $\hbar m$ along the propagation direction. 
To analyze how an OAM of the (free) electrons may influence the RR process and especially the properties of the emitted light, we use here the density matrix formalism based on the Schr\"odinger equation. 
Such a non-relativistic approach is well justified for (relatively) slow collisions with light ions and enables us to derive \textit{analytical} expressions for the angular distribution and polarization of the recombination photons. 
Based on these expressions, we show that both the emission pattern and the linear polarization of the RR radiation may change significantly if a twisted instead of a plane-wave electron beam is used. 
These modifications can be easily measured with present-day detectors, and the detector signal may serve as a ``feedback'' to control the twisted electron beam.

Hartree atomic units ($\hbar = e = m_e = 1$) are used throughout the paper.

\smallskip

%
%
% ----------------------------------------------- Theory part ------------------------------------------------------------------------------ %
%
%

\textit{Theoretical background}--- Photons emitted in the capture of electrons to an ionic bound state $\ket{b}$ are characterized by their energy, angular distribution and polarization. 
While the energy $E_\gamma = E_{\rm kin} + \bigl| E_b \bigr|$ is uniquely defined by the  velocity $v$ of the incident electrons (and, hence, by $E_{\rm kin} = v^2/2$) as well as the binding energy $E_b$, the angular and polarization properties of the RR radiation are sensitive to details of the electron-photon interaction. 
In atomic theory, one usually describes these properties by (i) the probability $W \big( {\bf k} \big)$ to emit a photon into a certain direction ${\bf k}/k$, ${\bf k}$ being the wave vector, and (ii) the three Stokes parameters $P_i$, $i=1,2,3$. 
The first two Stokes parameters quantify ``relative'' asymmetries between intensities $I_{\chi}$ of light that is \textit{linearly} polarized under different angles $\chi$ with respect to the reaction plane \cite{ReactionPlane}: $P_1 = (I_0 - I_{90})/(I_0 + I_{90})$ and $P_2 = (I_{45} - I_{135})/(I_{45} + I_{135})$. 
The parameter $P_3$ reflects the degree of \textit{circular} polarization of the emitted photons.

The emission probability $W \big( {\bf k} \big)$ and the Stokes parameters $P_i$ of the RR photons are described most conveniently within the density matrix framework \cite{Balashov2000,Blum1981}. 
Using this formalism, the aforementioned quantities are directly related to the so-called photon-spin density matrix
\begin{align}
   \label{Eq:SpinDensityMatrix}
   \mem{{\bf k} \lambda}{\hat{\rho}_\gamma }{{\bf k} \lambda^\prime}_{\lambda, \lambda^\prime = \pm 1} \! = \! \frac{W\bigl({\bf k}\bigr)}{2}
   \twobytwo{1 + P_3}{{-P_1} + \mathrm{i} \, P_2}{{-P_1} - \mathrm{i} \, P_2}{1 - P_3} \!,
\end{align}
where $\lambda = \pm 1$ is the helicity of the emitted photon, i.e. its spin projection onto the direction of propagation.

Before we derive the density matrix~(\ref{Eq:SpinDensityMatrix}) for the capture of twisted electrons, let us briefly recall the recombination process for plane-wave electrons. 
By restricting ourselves to the RR of non-relativistic electrons into the $1s$ ground state of bare ions with low nuclear charge $Z$, we can obtain the photon density matrix~(\ref{Eq:SpinDensityMatrix}) as
\begin{align}
 \label{Eq:SpinDensityMatrixPlaneWave}
 \bra{{\bf k} \lambda} \hat{\rho}_\gamma^{\rm pl} \ket{{\bf k} \lambda^\prime} = M^{\rm pl,\ast}_{{\bf k}, {\bf p}}(\lambda) \, M^{\rm pl}_{{\bf k}, {\bf p}}(\lambda^\prime) \, ,
\end{align}
where the (transition) amplitude
\begin{align}
 \label{Eq:TransitionAmplitudesPlaneWaveIntegral}
 M^{\rm pl}_{{\bf k}, {\bf p}}(\lambda) = \int \psi_{f}^{\ast}({\bf r}) \, \mathrm{e}^{-\mathrm{i} {\bf k} \cdot {\bf r}} ({\bf e}_\lambda^\ast \cdot \hat{\bf p}) \, \psi_{i}^{\rm pl}({\bf r}) \, {\rm d}^3 {\bf r}
\end{align}
describes the emission of a photon with polarization ${\bf e}_\lambda$ and wave vector ${\bf k}$.  
In this process, moreover, the electron performs a transition from the (initial) continuum state $\psi_i^{\rm pl}({\bf r}) \equiv \psi_{\bf p}^{\rm pl}({\bf r})$ with momentum ${\bf p}$ into the (final) ionic ground state $\psi_{f}({\bf r}) = \psi_{\rm 1s}({\bf r}) \equiv \sqrt{Z^3/\pi} \ \mathrm{e}^{-Z r}$, and couples to the photon via the momentum operator ${\bf \hat{p}} = - \mathrm{i} \, \partial/\partial {\bf r}$.
The amplitude~(\ref{Eq:TransitionAmplitudesPlaneWaveIntegral}) can be evaluated analytically to give
\begin{align}
 \label{Eq:TransitionAmplitudesPlaneWave}
  M^{\rm pl}_{{\bf k}, {\bf p}}(\lambda) = 8 \sqrt{\pi Z^5} \, \frac{({\bf e}_\lambda^\ast \cdot {\bf p})}{\left [\left({\bf p} - {\bf k} \right)^2 + Z^2 \right ]^2}
\end{align}
if the incoming electron is \textit{approximated} as a plane wave $\psi_{i}^{\rm pl}({\bf r}) = \mathrm{e}^{\mathrm{i} {\bf p} \cdot {\bf r}}$. 
Although the plane-wave approximation neglects the (Coulomb) interaction of the incoming electron with the ion, it is well justified for kinetic energies much higher than the ionization threshold \cite{Bethe1957}. 
Since in the present study we focus on energetic (but still non-relativistic) collisions, we will use Eq.~(\ref{Eq:TransitionAmplitudesPlaneWave}) in the calculations below.

Inserting the transition amplitude~(\ref{Eq:TransitionAmplitudesPlaneWave}) into Eqs.~(\ref{Eq:SpinDensityMatrix})--(\ref{Eq:SpinDensityMatrixPlaneWave}) leads (after some simple algebra) to the ``standard'' expressions of the angular distribution and polarization of the RR radiation that follows the capture of plane-wave electrons into the ground state of a (hydrogen-like) ion \cite{Bethe1957}.
More specifically, the emitted photons are completely linearly polarized within the reaction plane, $P_1 = 1$, and their emission pattern reads $W({\bf k}) \propto \sin^2\theta_k / \left[ p^2 + k^2 + Z^2 - 2 p k \cos\theta_k \right]^4$, where $\theta_k$ denotes the polar angle of the wave vector ${\bf k}$ with regard to the direction of the incident electron beam.

To explore how a twisted electron beam affects the angular and polarization properties of the recombination light, we return to the general expression~(\ref{Eq:SpinDensityMatrix}) for the photon-spin density matrix. 
Similar to the capture of plane-wave electrons, the evaluation of this matrix is traced back to the RR amplitudes and, hence, to the explicit form of the electron wave functions. 
While the final (bound) state is again described by the function $\psi_f({\bf r}) = \psi_{1s}({\bf r})$, the initial wave function $\psi_i({\bf r})$ should be modified to represent a twisted electron state. 
Following recent studies \cite{Bliokh2012,Ivanov2011,Schattschneider2011}, we assume here that the electrons are prepared as a (twisted) \textit{wave packet} with well-defined values of (i) the linear momentum $p_z$ and (ii) the OAM $m = 0, \pm 1, \pm 2 \, ...$ along the propagation direction, which is taken as the quantization $z$-axis. 
In this case, the initial-state wave function reads
\begin{align}
  \label{Eq:TwistedElectronWavefunction}
  \psi^{\rm tw}_i({\bf r}) \equiv \psi^{\rm tw}_{\varkappa m p_z}({\bf r}) = \int a_{\varkappa m}({\bf p}_\perp) \, \mathrm{e}^{\mathrm{i} \left({\bf p}_\perp \cdot {\bf r}_\perp + p_z z \right)} \, \frac{{\rm d}^2 {\bf p}_\perp}{(2\pi)^2} \, ,
\end{align}
where the (Fourier) coefficient $a_{\varkappa m}$ is
\begin{align}
  \label{Eq:Amplitude}
  a_{\varkappa m}({\bf p}_\perp) = \sqrt{\frac{2\pi}{p_\perp}} (-\mathrm{i})^m \mathrm{e}^{\mathrm{i} m \varphi_p} g_\varkappa(p_\perp) \, ,
\end{align}
and the function $g_\varkappa(p_\perp)$ describes the transverse momentum profile of the beam. 
We suppose here a Gaussian function $g_\varkappa(p_\perp) = (\pi \sigma_\varkappa^2)^{-1/4} \, \mathrm{e}^{-(p_\perp - \varkappa)^2/(2 \sigma_\varkappa^2)}$ with a width $\sigma_\varkappa \ll \varkappa$. 
This implies that the wave packet~(\ref{Eq:TwistedElectronWavefunction}) has a mean kinetic energy $E_{\rm kin} = \left(\varkappa^2 + p_z^2 \right) / 2$ and that its momentum distribution peaks sharply at $p_\perp = \varkappa$.

\begin{figure}[tb]
   \includegraphics[width= 6 cm,clip=true]{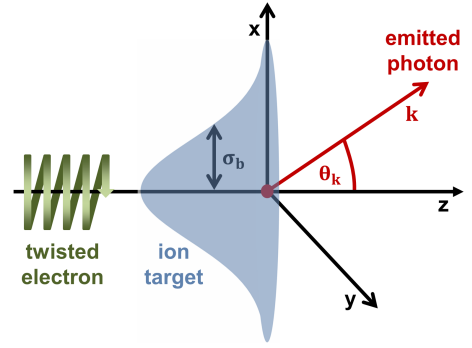}
   \caption{\label{Fig:Geometry} Geometry for the radiative capture of twisted electrons by bare ions. The (quantization) z-axis is chosen along the propagation direction of the electron wave. The recombination photon is emitted under a polar angle $\theta_k$. Its momentum ${\bf k}$ defines together with the z-axis the x-z-plane.}
\end{figure}

The twisted electron wave~(\ref{Eq:TwistedElectronWavefunction})--(\ref{Eq:Amplitude}) can be considered as a coherent superposition of plane waves $\mathrm{e}^{\mathrm{i} {\bf p} \cdot {\bf r}} = \mathrm{e}^{\mathrm{i} \left({\bf p}_\perp \cdot {\bf r}_\perp + p_z z \right)}$ whose momentum vectors ${\bf p}$ form the surface of a cone with (mean) opening angle $\theta_p = \arctan \left(\varkappa/p_z \right)$. 
As known from the literature \cite{Bliokh2012,Ivanov2011,Schattschneider2011,Matula2013}, it is the azimuthal phase factor $\exp(\mathrm{i} m \varphi_p)$ of the superposition coefficients~(\ref{Eq:Amplitude}) that gives rise to the well-defined (projection of) OAM $\hbar m$ and leads also to a pronounced \textit{spatial} profile of the electron packet. 
That is, the electron probability density exhibits for $m\neq0$ a concentric ring structure with a central dark spot within a plane perpendicular to the $z$-(propagation) axis.

Owing to the complex spatial structure of the state~(\ref{Eq:TwistedElectronWavefunction}), the properties of the emitted photons depend on the \textit{position} of a target ion within the electron wave front. 
This is taken into account by translating the final-state wave function 
\begin{align}
   \label{Eq:translation}
   \psi_f({\bf r} - {\bf b}_\perp) = \mathrm{e}^{-\mathrm{i} {\hat {\bf p}} \cdot {\bf b}_\perp } \psi_f({\bf r}) \, 
\end{align}
from the (zero-intensity) center of the electron wave by a (transverse) vector ${\bf b}_\perp$.
Employing expression~(\ref{Eq:translation}) and the initial-state wave function~(\ref{Eq:TwistedElectronWavefunction}), we finally obtain the amplitude
\begin{align}
   \label{TransitionAmplitudeTwistedWave}
   M_{{\bf k}, \varkappa, p_z, {\bf b}_\perp}^{\rm tw}(\lambda) &= \int \psi_f^\ast({\bf r} - {\bf b}_\perp) \, \mathrm{e}^{-\mathrm{i} {\bf k} \cdot {\bf r}} ({\bf e}_\lambda^\ast \cdot \hat{\bf p}) \, \psi_i^{\rm tw}({\bf r}) \, {\rm d}^3{\bf r} \nonumber \\
   &= \int \mathrm{e}^{\mathrm{i} ({{\bf p}_\perp - {\bf k}_\perp}) \cdot {\bf b}_\perp} \, a_{\varkappa m}({\bf p}_\perp) \,
   M_{{\bf k}, {\bf p}}^{\rm pl}(\lambda) \, \frac{{\rm d}^2 {\bf p}_\perp}{(2\pi)^2} \, 
\end{align}
that describes the capture of a twisted electron by an ion with impact parameter ${\bf b}_\perp$. 
As seen from the second line of Eq.~(\ref{TransitionAmplitudeTwistedWave}), such an amplitude can be represented in terms of the standard ``plane-wave'' matrix elements~(\ref{Eq:TransitionAmplitudesPlaneWave}).

The amplitudes~(\ref{TransitionAmplitudeTwistedWave}) provide the main building blocks to construct the photon density matrix for K-shell RR of twisted electrons. 
Beside the amplitudes, however, this matrix depends also on the set-up of a particular ``experiment''. 
Typically, the twisted electrons collide with some macroscopic ion target and, hence, no information is available about the impact parameter ${\bf b}_\perp$. 
The photon density matrix then reads
\begin{align}
 \label{Eq:DensityMatrixTwistedWave}
 \bra{{\bf k} \lambda} \hat{\rho}_\gamma^{\rm tw} \ket{{\bf k} \lambda^\prime} = &\int f({\bf b}_\perp) \, M^{\rm tw,\ast}_{{\bf k}, \varkappa, p_z, {\bf b}_\perp}(\lambda) \nonumber \\
 &\times M^{\rm tw}_{{\bf k}, \varkappa, p_z,{\bf b}_\perp}(\lambda^\prime) \, {\rm d}^2 {\bf b}_\perp,
\end{align}
where $f({\bf b}_\perp)$ denotes the probability to find a target ion at a (transverse) position ${\bf b}_\perp$ with respect to the center of the incident electron beam. 
By choosing a Gaussian function $f({\bf b}_\perp) = \left(2 \pi \sigma_b^2 \right)^{-1} \, \mathrm{e}^{-{\bf b}_\perp^2/(2\sigma_b^2)}$ with a width $\sigma_b$ for the distribution of the ions, c.f.~Fig.~\ref{Fig:Geometry}, and by making use of Eq.~(\ref{TransitionAmplitudeTwistedWave}), we find
\begin{align}
\begin{split}
 \label{Eq:IntermediateDensityMatrixTwistedWave}
 &\bra{{\bf k} \lambda} \hat{\rho}_\gamma^{\rm tw} \ket{{\bf k} \lambda^\prime} = \int \exp\left[-2 \varkappa^2 \sigma_b^2 \sin^2\left(\frac{\varphi_p^{} - \varphi_p^\prime}{2} \right) \right] \\[0.2cm]
 &\times a_{\varkappa m}^\ast({\bf p}_\perp^{}) \, a_{\varkappa m}({\bf p}_\perp^\prime) \, M_{{\bf k}, {\bf p}}^{\rm pl,\ast}(\lambda) \, M_{{\bf k}, {\bf p}^\prime}^{\rm pl}(\lambda^\prime) \, \frac{{\rm d}^2 {\bf p}_\perp^{}}{(2\pi)^2} \frac{{\rm d}^2 {\bf p}_\perp^\prime}{(2\pi)^2} \, 
\end{split}
\end{align}
after integrating over the impact parameter ${\bf b}_\perp$.

If---as usual in collision experiments---the width of the incoming electron beam is considerably smaller than the size of the ionic target and, hence, $1/\sigma_{\varkappa} \ll \sigma_b$, the density matrix can be evaluated analytically. 
In this case the exponent in Eq.~(\ref{Eq:IntermediateDensityMatrixTwistedWave}) acts (up to some factor) as a delta distribution $\delta(\varphi_p^{} - \varphi_p^\prime)$, thus allowing us to write
\begin{align}
  \label{Eq:FinalDensityMatrix}
  \bra{{\bf k} \lambda} \hat{\rho}_\gamma^{\rm tw} \ket{{\bf k} \lambda^\prime} &\propto \int M_{{\bf k}, {\bf p}}^{\rm pl,\ast}(\lambda) \, M_{{\bf k}, {\bf p}}^{\rm pl}(\lambda^\prime) \, {\rm d} \varphi_p \nonumber \\
  &\propto J_1({\bf k}) + \lambda \lambda^\prime J_2({\bf k}) \, .
\end{align}
The functions $J_1$ and $J_2$ do not depend on the polarization (helicity) states of the emitted radiation and can be expressed solely in terms of kinematic parameters such as the opening angle $\theta_p$ and the momentum $p=\sqrt{\varkappa^2+p_z^2}$ of the twisted electron packet and the wave vector ${\bf k}$ of the photon:
\begin{align}
   \label{Eq:AandB}
   J_1({\bf k}) &= \frac{1}{4 v^4 (1 - u^2)^{7/2}} \left ( 1 - u^2 \right ) \sin^2\theta_p, \\
   \begin{split}
   J_2({\bf k}) &= \frac{1}{8 v^4 (1 - u^2)^{7/2}} \Big[ 2 \left( 2 + 3 u^2 \right) \sin^2\theta_k \cos^2\theta_p \\[0.2cm]
   & \quad - \left( 4 u + u^3 \right) \sin2\theta_k \sin2\theta_p \\
   & \quad + 2 \left( 1 + 4 u^2 \right) \cos^2\theta_k \sin^2\theta_p \Big] \, ,
   \end{split}
\end{align}
where the notations
\begin{align}
\label{Eq:Abbreviations}
\begin{split}
  u &= \frac{2 p k}{v} \sin\theta_p \sin\theta_k, \ 0 \le u < 1, \\
 v &= p^2 + k^2 + Z^2 - 2 p k \cos\theta_p \cos\theta_k, \ 0 < v \,,
\end{split}
\end{align}
are used and the emission angle $\theta_k$ is defined with respect to the incident beam direction (cf. Fig.~\ref{Fig:Geometry}).

%\smallskip

By employing the photon-spin density matrix~(\ref{Eq:FinalDensityMatrix}) and Eq.~(\ref{Eq:SpinDensityMatrix}), one can study the angular and polarization properties of recombination photons for an (incident) twisted electron beam. 
For example, if we take the trace of~(\ref{Eq:FinalDensityMatrix}) over the photon helicities, we immediately find
\begin{align}
   \label{Eq:AngularDistribution}
    W\bigl({\bf k}\bigr) = \sum\limits_{\lambda} \bra{{\bf k} \lambda} \hat{\rho}_\gamma^{\rm tw} \ket{{\bf k} \lambda} = {\cal N}
    \left( J_1({\bf k}) + J_2({\bf k}) \right) \, ,
\end{align}
where ${\cal N} = 3 \left[(p^2-k^2)^2+ 2 \, Z^2 (p^2+k^2) + Z^4 \right]^2$ is defined from the normalization condition $\int W({\bf k}) {\rm d}\Omega_k = 4 \pi$. 
The polarization of the RR radiation is characterized, moreover, by the single Stokes parameter
\begin{align}
 \label{Eq:P1}
 P_1 = \frac{J_2({\bf k}) - J_1({\bf k})}{J_1({\bf k}) + J_2({\bf k})} \, ,
\end{align}
while $P_2$ and $P_3$ vanish identically.

Eqs.~(\ref{Eq:AngularDistribution}) and~(\ref{Eq:P1}) describe the emission pattern and the Stokes parameter $P_1$ of the RR radiation for the twisted electron waves~(\ref{Eq:TwistedElectronWavefunction}) with any arbitrary ratio of the transverse and longitudinal momenta $\varkappa$ and $p_z$. 
Most studies in the past, however, have dealt with \textit{paraxial} beams, characterized by small ratios $\varkappa / p_z \ll 1$ and, hence, opening angles $\theta_p \lesssim 1^\circ$ \cite{Uchida2010,Verbeeck2010,McMorran2011,Verbeeck2012,Saitoh2012}. 
For such beams, Eqs.~(\ref{Eq:AandB})--(\ref{Eq:P1}) can be further simplified to give
\begin{align}
\label{Eq:paraxial_angular}
  W({\bf k}) &\simeq \frac{\cal N}{2} \, \frac{\sin^2\theta_k + \sin^2\theta_p \, P_2(\cos\theta_k)}{\left( p^2 + k^2 + Z^2 - 2 p \, k \cos\theta_k \right)^4} \, , \\ 
\label{Eq:paraxial_P1}
  P_1 &\simeq 1 - \frac{\sin^2\theta_p}{\sin^2\theta_k + \sin^2\theta_p \, P_2(\cos\theta_k)},
\end{align}
where $P_2(\cos\theta_k) = (3\cos^2\theta_k - 1)/2$ denotes the second-order Legendre polynomial. 
As seen from these formulas, both the angular distribution and the linear polarization of the emitted photons are described by a sum of the ``standard'' expression, as obtained for the plane-wave electrons, \textit{and} an additional term proportional to $\sin^2\theta_p$. 
This latter term arises due to the twisted nature of the incident electron and, hence, disappears when the initial wave function~(\ref{Eq:TwistedElectronWavefunction}) is reduced to the plane-wave solution at $\theta_p \to 0$.

\smallskip

%
%
% ----------------------------------------------- Results and discussion  ------------------------------------------------------------- %
%
%

%
%
%
\begin{figure}[tb]
   \includegraphics[width=8.6 cm,clip=true]{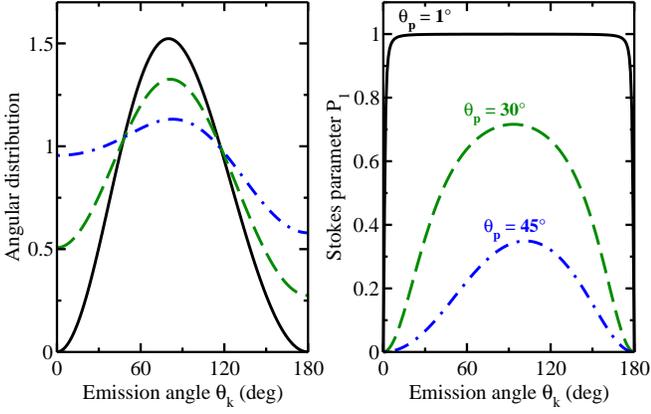}
   \caption{\label{Fig2} Angular distribution (left panel) and Stokes parameter $P_1$ (right panel) of photons emitted in the capture of twisted electrons into the $1s$ ground state of a hydrogen atom. Calculations, based on Eqs.~(\ref{Eq:AngularDistribution}) and~(\ref{Eq:P1}), were performed for an electron beam with an energy of $E_{\rm kin} = 2 \ {\rm keV}$ and for three different values of the opening angle: $\theta_p = 1^\circ$ (black solid lines), $\theta_p = 30^\circ$ (green dashed lines) and $\theta_p = 45^\circ$ (blue dashed-dotted lines).}
\end{figure}

\textit{Results and discussion}--- Both the general and the approximate expressions~(\ref{Eq:AngularDistribution})--(\ref{Eq:P1}) and~(\ref{Eq:paraxial_angular})--(\ref{Eq:paraxial_P1}) for the angular distribution and linear polarization of the RR photons were derived for twisted electron beams whose width $\sim 1/\sigma_{\varkappa}$ is much smaller compared to the size of a \textit{macroscopic} target $\sigma_b$. 
For such a scenario, which can be easily realized experimentally, the properties of the emitted photons are insensitive to the topological charge $m$ and depend only on the (mean) transverse momentum $\varkappa$ of the electron packet as characterized by the angle $\theta_p$. 
In Fig.~\ref{Fig2} we display the angular distribution (left panel) and the Stokes parameter $P_1$ (right panel) as resulting for the radiative capture of twisted electrons with $E_{\rm kin} = 2$~keV into the ground state of (finally) hydrogen atoms, and for different opening angles $\theta_p$. 
As expected already from Eqs.~(\ref{Eq:paraxial_angular})--(\ref{Eq:paraxial_P1}), both (angular and polarization) properties resemble the ``plane-wave'' behaviour $W\bigl({\bf k}\bigr) \sim \sin^2\theta_k$ and $P_1 \sim 1$ in the {\it paraxial} regime where $\theta_p \lesssim 1^\circ$. 
However, a strong enhancement of the RR-photon yield can be observed in the forward ($\theta_k \lesssim 30^\circ$) and backward ($\theta_k \gtrsim 150^\circ$) directions with respect to the incident electron beam as the opening angle is increased. 
At these emission angles, moreover, the Stokes parameter $P_1$ drops by factors of 1.5 to 10 if $\theta_p$ changes from $1^\circ$ to $45^\circ$. 
By employing modern detectors such a modification of the polarization (as well as of the angular distribution) can be easily observed, and thus these measurements may be used to further explore the interaction of twisted electrons with matter.

\begin{figure}[tb]
  \includegraphics[width=8.6 cm,clip=true]{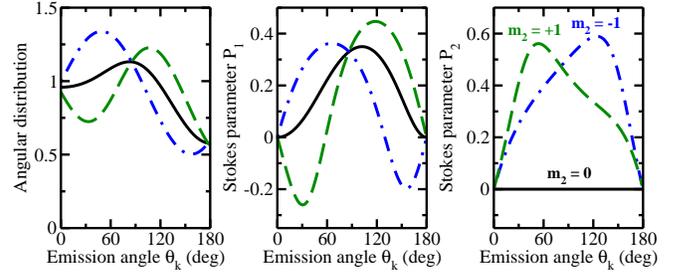}
  \caption{\label{Fig3} Angular distribution (left panel) and Stokes parameters $P_1$ (middle panel) and $P_2$ (right panel) of photons emitted in the capture of twisted electrons into the $1s$ ground state of a hydrogen atom. Here the electrons are in a superposition two states with relative phase $\zeta=\pi/4$ and projections of the OAM $m_1 = 0$, and $m_2 = 0$ (black solid lines), $m_2 = +1$ (green dashed lines) or $m_2 = -1$ (blue dashed-dotted lines), respectively. For $m_1=m_2=0$, this corresponds to a beam in the pure state with zero OAM. Results are shown for an electron energy $E_{\rm kin} = 2 \ {\rm keV}$ and within the non-paraxial regime~$\theta_p = 45^\circ$.}
\end{figure}

Until now, we have discussed the capture of twisted electrons that are prepared in a pure state with well-defined OAM $\hbar m$. 
As mentioned already, in such a case the angular and polarization properties of the emitted light are sensitive to the opening angle $\theta_p$, but not to the topological charge $m$ of the beam. 
This insensitivity can be overcome if the incident electron packet is prepared as a coherent superposition of two states with different projections $m_1$ and $m_2$ of orbital momentum onto the mutual propagation $z$-axis \cite{Guzzinati2013,Ivanov2012}:
\begin{align}
   \label{eq:superposition_twisted}
   \psi_i^{\rm tw}({\bf r}) = c_1 \, \psi_{\varkappa m_1 p_z}^{\rm tw}({\bf r}) + c_2 \, \mathrm{e}^{\mathrm{i} \zeta} \, \psi_{\varkappa m_2 p_z}^{\rm tw}({\bf r}) \, ,
\end{align}
where $c_1$ and $c_2$ are (real) weight factors with $|c_1|^2 + |c_2|^2 = 1$ and $\zeta$ is a relative phase. 
By inserting this formula into Eqs.~(\ref{TransitionAmplitudeTwistedWave})--(\ref{Eq:DensityMatrixTwistedWave}), the photon density matrix 
\begin{align}
\label{eq:density_matrix_superposition}
\begin{split}
& \bra{{\bf k} \lambda} \hat{\rho}_\gamma^{\rm tw} \ket{{\bf k} \lambda^\prime} \propto \int M_{{\bf k}, {\bf p}}^{{\rm pl},\ast}(\lambda) \, M_{{\bf k}, {\bf p}}^{\rm pl}(\lambda^\prime) \\
& \quad \times \bigl \{ 1 + 2 |c_1 c_2| \cos\left[{\Delta m} (\varphi_p - \pi/2) + \zeta \right]\bigr\} \, {\rm d} \varphi_p,
\end{split}
\end{align}
can be written as a sum of the matrix~(\ref{Eq:FinalDensityMatrix}) that was obtained for the individual electron beams as well as an interference term that depends on the difference $\Delta m = m_2 - m_1$. 
Such a $\Delta m$-dependence translates directly to the properties of the emitted light. 
As seen from Fig.~\ref{Fig3}, for example, both the angular distribution and the polarization $P_1$ are strongly affected by the variation of $\Delta m$ if the electron packets are prepared in a state with energy $E_{\rm kin} = 2$~keV, opening angle $\theta_p = 45^\circ$, and OAM $m_1 = 0$ and $m_2 = 0, \pm 1$. 
Moreover, the second Stokes parameter $P_2$, which is identically zero for a single wave packet~(\ref{Eq:TwistedElectronWavefunction}) with a well-defined $m$, becomes considerably large if $\Delta m \ne 0$ (cf. right panel of Fig.~\ref{Fig3}). 
We therefore conclude that the $\Delta m$-dependence enables one to \textit{determine} the OAM of an electron packet when it is mixed coherently with a ``reference'' beam whose topological charge $m$ is well known.
 
\smallskip

%
%
% ----------------------------------------------- Conclusions ------------------------------------------------------------- %
%
%

\textit{Conclusions}--- In summary, we have performed a theoretical study of the radiative recombination of twisted electrons with low-$Z$ bare ions. 
Based on solutions of the non-relativistic Schr\"odinger equation for the capture into the ionic ground state, analytical expressions were derived for (i) the angular distribution and (ii) the Stokes parameters of the emitted photons. 
Depending on the particular setup of the recombination experiment, these (angular and polarization) properties appear to be sensitive to the kinematic parameters such as the ratio $\varkappa/p_z$ of transverse to longitudinal momentum, as well as to the ``twistedness'' of the incident electron beam. 
We argue, therefore, that observation of the K-shell radiative recombination of twisted electrons will not only provide more details about the fundamental light-matter interaction process but can also be used for the control of twisted electron beams.

\smallskip

%
%
% ------------------------------------------------- Acknowledgments --------------------------------------------
%
%

\begin{acknowledgments}
O.~M. and A.~S. acknowledge support from the Helmholtz Gemeinschaft and GSI (Nachwuchsgruppe VH-NG-421). 
A.~G.~H. acknowledges support from the GSI Helmholtzzentrum and the University of Heidelberg. 
V.~G.~S. is supported by the Russian Foundation for Basic Research via grants 13-02-00695 and NSh-3802.2012.2.
\end{acknowledgments}

%
%
% ------------------------------------------------- References --------------------------------------------
%
%


\begin{thebibliography}{99}
\bibitem{ReactionPlane} The reaction plane is formed by the directions of the incident electron and the emitted photon.
\end{thebibliography}
\end{document}